\documentclass{emulateapj}

\begin{document}

\shortauthors{Luhman et al.}
\shorttitle{T Dwarf Companions}

\title{Discovery of Two T Dwarf Companions with the Spitzer Space Telescope}

\author{K. L. Luhman\altaffilmark{1},
B. M. Patten\altaffilmark{2},
M. Marengo\altaffilmark{2},
M. T. Schuster\altaffilmark{2,3},
J. L. Hora\altaffilmark{2},
R. G. Ellis\altaffilmark{2},
J. R. Stauffer\altaffilmark{4},
S. M. Sonnett\altaffilmark{2,5},
E. Winston\altaffilmark{2,6},
R. A. Gutermuth\altaffilmark{2,6},
S. T. Megeath\altaffilmark{7},
D. E. Backman\altaffilmark{8},
T. J. Henry\altaffilmark{9},
M. W. Werner\altaffilmark{10},
and G. G. Fazio\altaffilmark{2}}

\altaffiltext{1}{Department of Astronomy and Astrophysics,
The Pennsylvania State University, University Park, PA 16802;
kluhman@astro.psu.edu.}

\altaffiltext{2}{Harvard-Smithsonian Center for Astrophysics, 60 Garden St.,
Cambridge, MA 02138.}

\altaffiltext{3}{School of Physics and Astronomy, University of Minnesota,
Minneapolis, MN 55455.}

\altaffiltext{4}{Spitzer Science Center, Caltech MS 314-6, Pasadena, CA 91125.}

\altaffiltext{5}{Institute for Astronomy, The University of Hawaii at Manoa, 
2680 Woodlawn Drive, Honolulu, HI 96822.}

\altaffiltext{6}{Visiting Astronomer at the Infrared Telescope Facility, which 
is operated by the University of Hawaii under Cooperative Agreement no. NCC 
5-538 with the National Aeronautics and Space Administration, Office of Space 
Science, Planetary Astronomy Program.}

\altaffiltext{7}{Department of Physics and Astronomy, The University of Toledo,
2801 West Bancroft Street, Toledo, OH 43606.}

\altaffiltext{8}{SOFIA/SETI Institute, Mountain View, CA 94043.}

\altaffiltext{9}{Department of Physics and Astronomy, Georgia State University, 
Atlanta, GA 30302.}

\altaffiltext{10}{Jet Propulsion Laboratory, 4800 Oak Grove Drive, Pasadena, CA 
91109.}

\begin{abstract}

We report the discovery of T dwarf companions to the nearby stars HN~Peg 
(G0V, 18.4~pc, $\tau\sim0.3$~Gyr) and HD~3651 (K0V, 11.1~pc, $\tau\sim7$~Gyr). 
During an ongoing survey of $5\arcmin\times5\arcmin$ fields surrounding 
stars in the solar neighborhood with the Infrared Array Camera aboard 
the {\it Spitzer Space Telescope}, we identified these companions as 
candidate T dwarfs based on their mid-infrared colors.
Using near-infrared spectra obtained with SpeX at the NASA Infrared Telescope 
Facility, we confirm the presence of methane absorption that characterizes 
T dwarfs and measure spectral types of T2.5$\pm$0.5 and T7.5$\pm$0.5
for HN~Peg~B and HD~3651~B, respectively. By comparing our {\it Spitzer} data
to images from the Two-Micron All-Sky Survey obtained several years earlier, 
we find that the proper motions of HN~Peg~B and HD~3651~B are consistent 
with those of the primaries, confirming their companionship.
HN~Peg~B and HD~3651~B have angular separations of $43\farcs2$ and $42\farcs9$ 
from their primaries, which correspond to projected physical separations
of 795 and 476~AU, respectively.
A comparison of their luminosities to the values predicted by theoretical
evolutionary models implies masses of $0.021\pm0.009$ and 
$0.051\pm0.014$~$M_\odot$ for HN~Peg~B and HD~3651~B.
In addition, the models imply an effective temperature for HN~Peg~B that
is significantly lower than the values derived for other T dwarfs at
similar spectral types, which is the same behavior reported by Metchev
\& Hillenbrand for the young late-L dwarf HD~203030~B.
Thus, the temperature of the L/T transition appears to depend on surface 
gravity. Meanwhile, HD~3651~B is the first 
substellar companion directly imaged around a star that is known to harbor
a close-in planet from radial velocity surveys. 
The discovery of this companion supports the notion that the high 
eccentricities of close-in planets like the one near HD~3651 may be
the result of perturbations by low-mass companions at wide separations.

\end{abstract}


\section{Introduction}
\label{sec:intro}

Direct imaging surveys for resolved substellar companions are vital for
studies of brown dwarfs and planets. These surveys provide measurements 
of the frequency of companions as a function of various physical properties,
such as primary mass, companion mass, separation, and age, which in turn
can be used to constrain the mechanisms by which substellar companions form.
In addition, compared to their isolated counterparts, 
each newly uncovered companion represents a superior laboratory for studying 
the atmospheres and evolution of substellar objects 
because the age, distance, and metallicity of the primary, and 
hence the companion, are relatively easy to measure. 

A variety of criteria can be adopted for selecting primaries to search for 
companions.
Stars for which planets have been discovered through radial velocity monitoring 
and young nearby stars are particularly attractive categories to consider. 
Several direct imaging surveys have focused on known extrasolar planetary
systems \citep{luh02,pat02,mug05b,chau06}, producing one companion that
was initially described as a brown dwarf \citep{els01} but later shown to
be a white dwarf \citep{mug05a,lag06}.
Similar observations have been performed toward
young stars and brown dwarfs within 100~pc of the Sun
\citep[$\tau\sim10$-400~Myr,][]{cha03,neu00a,mz04,low05}, 
resulting in the discovery of several substellar companions that include 
G196-3~B \citep{reb98}, TWA~5~B \citep{low99,neu00b},
HR~7329~B \citep{low00,gue01}, AB~Pic~B \citep{cha05b},
HD~49197~B \citep{mh04}, 2M~1207-3932~B \citep{cha04,cha05a}, 
and HD~203030~B \citep{mh06}.
The spectral types of these companions range from late M through L
but do not extend into the cooler T types. 
The only known T dwarf companions to stars
consist of Gl~229~B \citep{nak95,opp95}, Gl~570~D \citep{bur00}, 
Gl~337~D \citep{wil01,bur05}, $\epsilon$~Ind~B and C \citep{sch03,mcc04}, 
and SCR~1845~B \citep{bil06}, all of which orbit 
nearby stars that are older than $\sim1$~Gyr and have no known planets.

Most resolved substellar companions have been identified through 
deep, high-resolution imaging at small separations ($\rho\lesssim10\arcsec$) 
with adaptive optics or the {\it Hubble Space Telescope} or through
shallow imaging at arbitrarily large separations with the Two-Micron 
All-Sky Survey (2MASS) or the Sloan Digital Sky Survey (SDSS). 
Because of its wide field of view ($\rho=2\farcm5$) and unprecedented 
sensitivity at mid-infrared (IR) wavelengths, 
the Infrared Array Camera \citep[IRAC;][]{faz04} 
aboard the {\it Spitzer Space Telescope} \citep{wer04} complements these
previous search methods by providing the best available sensitivity to 
low-mass companions at separations of $\rho\sim10$-$100\arcsec$
($r\sim100$-1000~AU) from nearby stars. 
To take advantage of these capabilities, 
we have obtained IRAC images of nearby young stars
and planet-bearing stars identified in radial velocity surveys.
In this paper, we report the discovery of two T dwarf companions to stars
in these samples, HN~Peg ($d=18.4$~pc, G0V) and HD~3651 ($d=11.1$~pc, K0V)
\footnote{In a paper submitted after this one, \citet{mug06} reported an
independent discovery of the companion to HD~3651.}.
The former is a young solar analog \citep{gai98} while the latter has a 
sub-Saturn planetary companion at 0.3~AU \citep{fis03,fis05}.
We begin by describing the IRAC images that were used to identify the T dwarf 
companions to these stars (\S~\ref{sec:phot}) and followup 
near-IR photometry and spectroscopy (\S~\ref{sec:spec}).
We then assess the evidence of companionship (\S~\ref{sec:bin}), 
measure spectral types (\S~\ref{sec:class}), and estimate the physical 
properties of these companions (\S~\ref{sec:prop}). Finally, we discuss
these newly discovered T dwarfs in the context of previously known 
brown dwarfs (\S~\ref{sec:disc}).

\section{Observations}
\label{sec:obs}

\subsection{Spitzer Mid-IR Imaging}
\label{sec:phot}

As a part of the Guaranteed Time Observations of the IRAC instrument team,
we obtained images at 3.6, 4.5, 5.8, and 8.0~\micron\ with IRAC aboard 
the {\it Spitzer Space Telescope} of 73 nearby young stars (PID=34) 
and 48 known extrasolar planetary systems (PID=48) between late 2003 and
early 2005. The plate scale and field of view of IRAC are $1\farcs2$ and
$5\farcm2\times5\farcm2$, respectively. The camera produces images with
FWHM$=1\farcs6$-$1\farcs9$ from 3.6 to 8.0~\micron\ \citep{faz04}.
The observing strategy and image reduction procedures were the same as those 
described by \citet{pat06a} for IRAC observations of M, L, and T dwarfs,
except for the following modifications \citep{pat06b}.
The data were processed using the software suite ``IRACproc" developed for 
this program to facilitate the combining of dithered or mapped IRAC data
and the rejection of cosmic-rays \citep{sch06}.
In order to obtain the best possible photometry of  faint
sources, we removed the light from each primary by fitting a scaled
IRAC point spread function in each band that was derived from IRAC 
observations of Vega, $\epsilon$~Eri and $\epsilon$~Ind~A \citep{mar06}.

\begin{figure}
\plotone{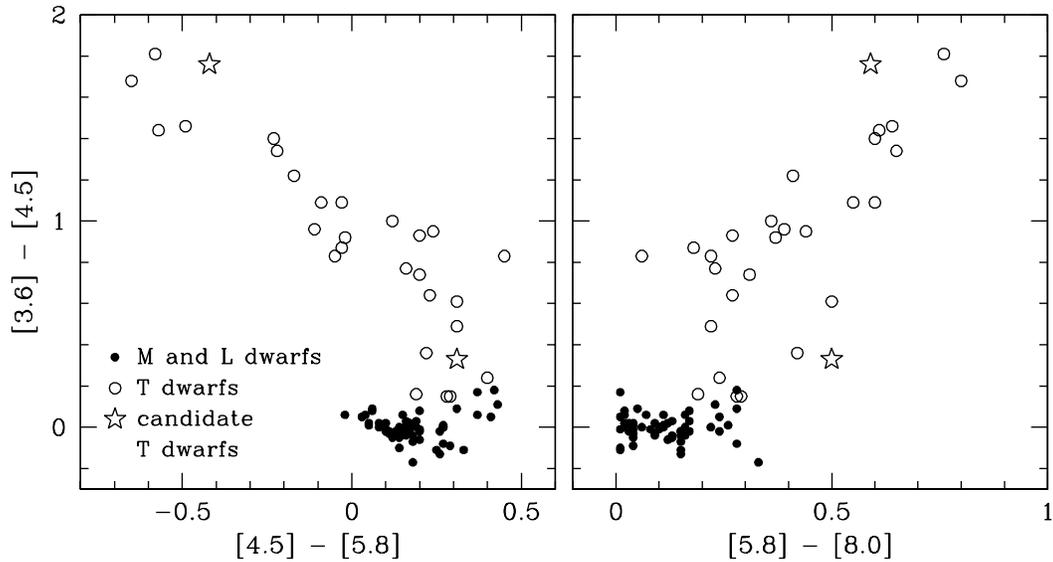}
\caption{ 
IRAC color-color diagrams for cool field dwarfs 
\citep[{\it open and filled circles},][]{pat06a} and two candidate companions 
of HN~Peg ({\it lower star}) and HD~3651 ({\it upper star}). 
The colors of these candidates are indicative of T dwarfs. 
}
\label{fig:cc}
\end{figure}

We measured photometry and astrometry of all point sources appearing in
the reduced images of each star. We then identified candidate T dwarf
companions by comparing these data to the colors of known T dwarfs 
\citep{pat06a}, which are highly distinctive from those of other astronomical 
sources. We did not search for M and L dwarf companions because their IRAC 
colors do not differ significantly from those of stars at earlier types. 
We also considered only candidates that had magnitudes that were 
consistent with the absolute magnitudes of known T dwarfs when placed at
the distance of the primary. These selection criteria were applied with a  
statistical technique based on the  
k-Nearest Neighbor \citep[$k$-NN,][]{fix51}, which is a  
nonparametric classifier appropriate when a physical  
model is not available or is not reliable enough for fitting the data. 
We adapted this data mining technique, commonly used in pattern  
recognition and unsupervised machine learning, to the case of  
astronomical photometric data. By defining an appropriate metric in  
the color and magnitude multi-dimensional space, we determined a  
``score" for each object, based on its ``distance" from the colors and  
magnitudes of known T dwarfs. A full description of our technique is 
presented in Marengo \& Sanchez (in preparation).
The two candidates with the best scores appeared in the images of HN~Peg
and HD~3651. Their positions in the IRAC color-color and color-magnitude 
diagrams are indicated Figs.~\ref{fig:cc} and \ref{fig:cmd} and their
photometric measurements are listed in Table~\ref{tab:data}. 
No other objects had scores that were comparable the ones for these two
candidates.
Followup observations of these candidates are described in the next section.

During a survey for debris disks with the Multiband Imaging Photometer for 
{\it Spitzer}, \citet{bry06} obtained 24~\micron\ images of HN~Peg. These
data encompass the candidate companion to this star that we have found with 
IRAC.  According to analysis that was kindly performed by G. Bryden, 
the candidate is not detected in those 24~\micron\ data.
Extrapolating from the IRAC data with a Rayleigh-Jeans distribution,
the expected 24~\micron\ flux from the candidate is $\sim0.1$~mJy while
the noise in the 24~\micron\ image is about twice this value. 
These relatively shallow observations were designed to detect an excess
above the stellar photosphere of HN~Peg rather than a faint companion.

\subsection{IRTF Near-IR Imaging and Spectroscopy}
\label{sec:spec}

We obtained near-IR spectra of the candidate T dwarf companions to HN~Peg 
and HD~3651 with the spectrometer SpeX \citep{ray03} at the
NASA Infrared Telescope Facility (IRTF) on the nights of 2006 June 15 and 30.
The instrument was operated in the prism mode with a $0\farcs8$ slit,
producing a wavelength coverage of 0.8-2.5~\micron\ and a resolution of
$R\sim100$.  For each candidate, we obtained 20 exposures distributed
between two positions along the slit. The individual exposure times were 
1.5~min, resulting in a total exposure time of 30~min for each object.
The slit was aligned to the parallactic angle for all observations.
The spectra were reduced with the Spextool package \citep{cus04} and
corrected for telluric absorption \citep{vac03}.

We also used the slit-viewing camera on SpeX to measure near-IR photometry 
for the candidate companions on the night of 2006 July 2.
The camera contained a $512\times512$ Aladdin 2 InSb array and had a 
plate scale of $0\farcs12$~pixel$^{-1}$ \citep{ray03}.
Imaging was performed through Mauna Kea Observatory (MKO) $J$, $H$, and $K$ 
filters \citep{sim02,tok02,tok05}.
In each filter, we obtained one 30~s exposure at each of nine dithered
positions. For the $K$-band imaging of the candidate companion to HD~3651,
we performed this cycle twice. 
Images for a given candidate and filter were median combined to produce a 
flat field image, which was then divided into the original exposures. 
Point sources in these images exhibited FWHM$=0\farcs8$.  
We extracted photometry for each candidate companion with the task PHOT under 
the IRAF package APPHOT using a radius of eight pixels. 
These data were calibrated with images of standard stars for the 
MKO system \citep{leg06}.
The $J$, $H$, and $K$ measurements for the two candidate companions are 
presented in Table~\ref{tab:data}.

\section{Analysis}
\label{sec:analysis}

\subsection{Evidence of Binarity}
\label{sec:bin}

When a candidate companion is identified in direct imaging of a star
in the solar neighborhood, the standard method of definitively assessing 
its status as a binary companion consists of checking whether it shares 
the same proper motion as the primary. 
To apply this method to the two candidate 
T dwarfs from our IRAC images of HN~Peg and HD~3651,
we needed an image of each object obtained at a second epoch. 
The candidate companion to HN~Peg was detected in each of the three 2MASS bands
and appears in the 2MASS Point Source Catalog. 
The candidate near HD~3651 is not in the 2MASS Point Source Catalog, but
through visual inspection of the images, we find that it was detected 
in $J$ and $H$. 

\begin{figure}
\plotone{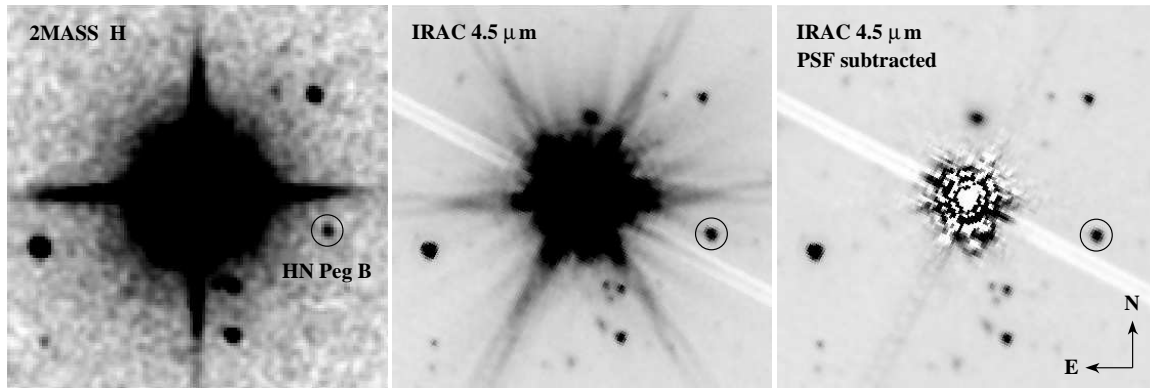}
\caption{
2MASS and IRAC images of HN~Peg~A and B. 
The size of each image is $2\arcmin\times2\arcmin$.
}
\label{fig:hnpeg}
\end{figure}

\begin{figure}
\plotone{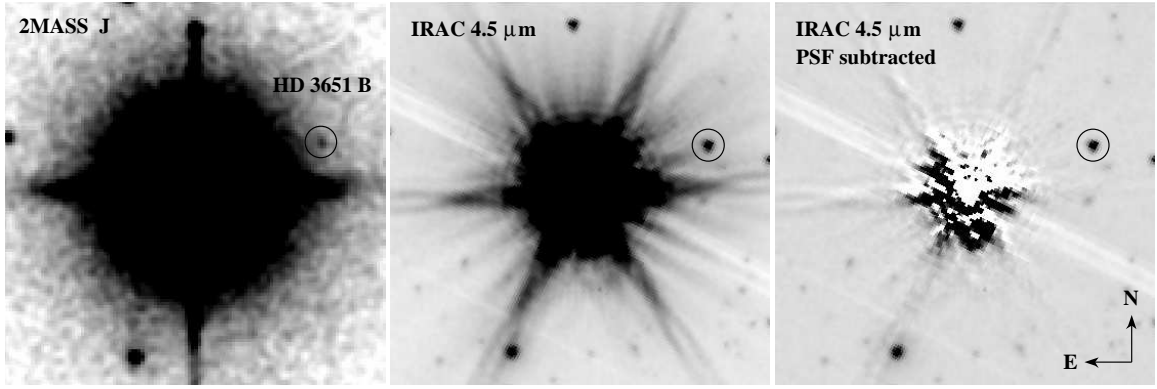}
\caption{ 
2MASS and IRAC images of HD~3651~A and B. 
The size of each image is $2\arcmin\times2\arcmin$.
}
\label{fig:hd3651}
\end{figure}

To measure the proper motion of each candidate, 
we identified sources that appear in both the IRAC and the 2MASS images
and measured their right ascensions and declinations in the 4.5~\micron\ image
of IRAC and in the 2MASS band in which the candidate companion has the highest 
signal-to-noise ratio ($H$ for HN~Peg and $J$ for HD~3651). The differences 
between the IRAC and 2MASS coordinates for these objects are plotted
in Figs.~\ref{fig:pm1} and \ref{fig:pm2}. 
The relatively small number of stars in each diagram is a reflection of the
requirement of a detection by 2MASS; the numbers of sources detected by the
IRAC images are much larger.  To correct for a small offset
between the coordinate systems of the IRAC and 2MASS images, we applied 
constant shifts to the right ascension and declination differences so 
that their average values were zero. In other words, we assume that the
average proper motion of stars surrounding each star is zero.
For comparison, we include in Figs.~\ref{fig:pm1} and \ref{fig:pm2} the
expected displacements of HN~Peg and HD~3651 during the time interval between
the 2MASS and IRAC observations ($\Delta\tau=5.7$ and 6.8~years) based
on their known proper motions \citep{per97}. 
In each of the two diagrams, one source exhibits a motion that is 
significantly non-zero and that is consistent with the expected motion 
of the primary. These two objects are the candidate T dwarfs that we 
found with IRAC. 
Thus, these data demonstrate that both candidates have common proper 
motions with the primaries, which conclusively establishes their companionship.
Hereafter in this paper, we refer to these new companions as HN~Peg~B and 
HD~3651~B\footnote{The planetary companion to HD~3651 from \citet{fis03} is 
designated with a lowercase ``b" based on the naming convention for planets 
from radial velocity surveys.}.
The positions of HN~Peg~B and HD~3651~B relative to their primaries are 
listed in Table~\ref{tab:data}. 2MASS and IRAC images of these systems
are shown in Figs.~\ref{fig:hnpeg} and \ref{fig:hd3651}.

\subsection{Spectral Classification}
\label{sec:class}

\begin{figure}
\plotone{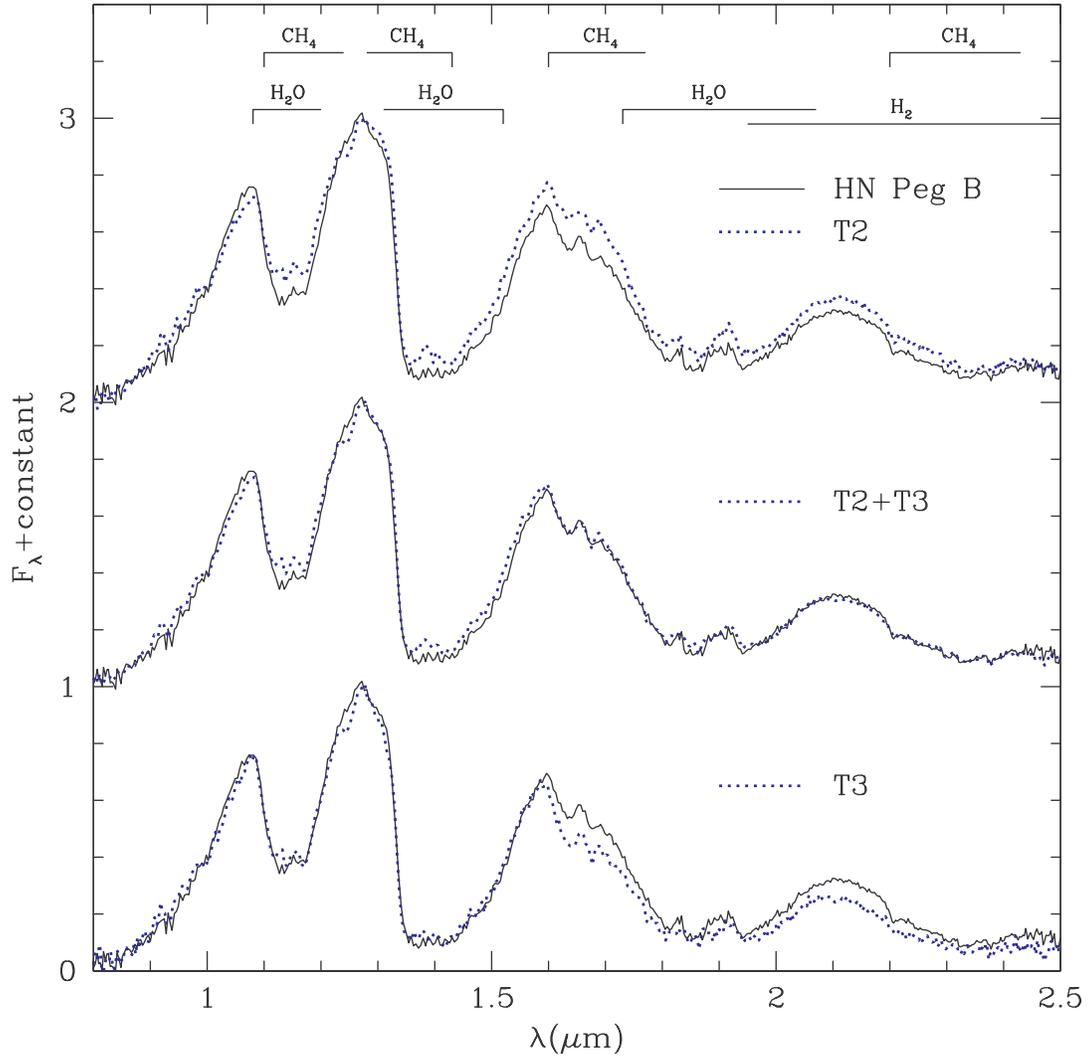}
\caption{
Near-IR spectrum of HN~Peg~B ({\it solid line}) compared to data for the 
standard T dwarfs SDSS~1254-0122 (T2) and 2MASS~1209-1004 (T3) from 
\citet[][{\it dotted lines}]{bur06b}. The average of these standard spectra 
provides the best match to the spectrum of HN~Peg~B.
The spectra have a resolution of $R=100$ and are normalized at 1.27~\micron.}
\label{fig:spec1}
\end{figure}

\begin{figure}
\plotone{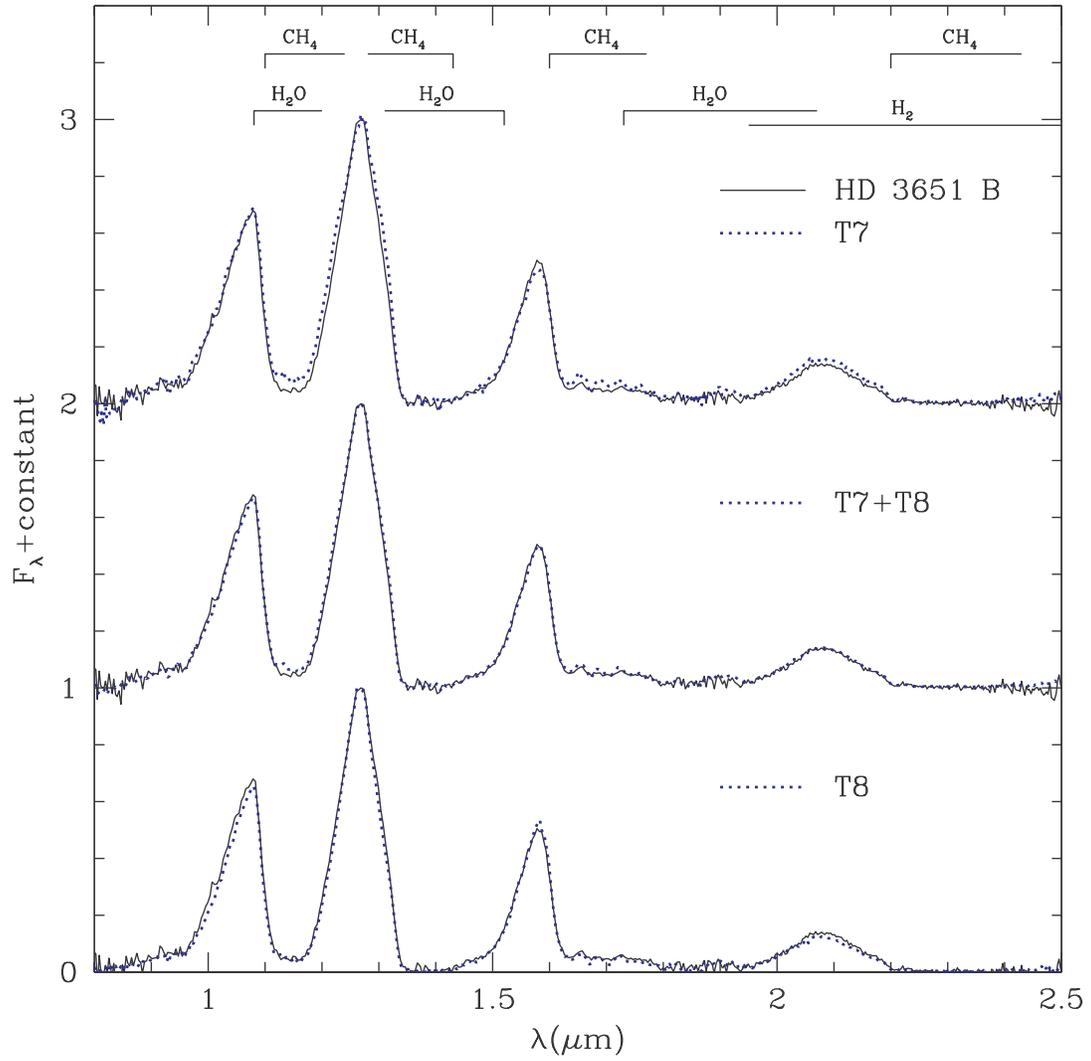}
\caption{
Near-IR spectrum of HD~3651~B ({\it solid line}) compared to data for the 
standard T dwarfs 2MASS~0727+1710 (T7) and 2MASS~0415-0935 (T8) from 
\citet[][{\it dotted lines}]{bur06b}. The average of these standard spectra 
provides the best match to the spectrum of HD~3651~B.
The spectra have a resolution of $R=100$ and are normalized at 1.27~\micron.}
\label{fig:spec2}
\end{figure}

In \S~\ref{sec:phot}, we identified HN~Peg~B and HD~3651~B as possible
T dwarfs based on their mid-IR colors and magnitudes. 
As shown in Figs.~\ref{fig:spec1} and \ref{fig:spec2}, our near-IR spectra of
both companions exhibit methane absorption, which is the
defining spectral feature of T dwarfs \citep{kir05}. 
To measure the precise spectral type of each companion, we visually compared
these spectra to the SpeX data for T dwarf standards that were presented by
\citet{bur06b}. In Figure~\ref{fig:spec1}, 
we find that the spectrum of HN~Peg~B agrees well with the average of the 
spectra of SDSS~1254-0122 (T2) and 2MASS~1209-1004 (T3). The match is 
clearly worse with either of these standards alone. Therefore, we assign a 
spectral type of T2.5$\pm0.5$ to HN~Peg~B. 

Although the overall spectrum of HN~Peg~B is fit well with standard T dwarfs, 
the K~I absorption lines at 1.25~\micron\ appear to be weaker in HN~Peg~B 
than in the standards. 
In comparison, these lines become weaker with lower gravity for late-M and 
early-L dwarfs \citep{mc04,kir06} but grow stronger with lower gravity 
at late-T types \citep{kna04}.
Based on the apparent youth of its primary (\S~\ref{sec:age}), HN~Peg~B should
have a lower surface gravity than the standard T dwarfs in 
Figure~\ref{fig:spec1}. 
Thus, these data for HN~Peg~B seem to indicate that the 
gravity dependence of the K~I lines is the same in early T dwarfs
as it is in M and L dwarfs. 
However, because the spectral resolution of our data is too low for reliable 
measurements of these lines, a spectrum at higher resolution is needed 
to explore this issue definitively.

In Figure~\ref{fig:spec2}, we compare HD~3651~B to 2MASS~0727+1710 (T7) 
and 2MASS~0415-0935 (T8) and the average of the two standards.
The differences between HD~3651~B and each of these three spectra are 
subtle, but close examination of each comparison indicates that 
the average of T7 and T8 provides the best match to HD~3651. 
For instance, the width of the continuum at 1.2-1.3~\micron, the depth of
the absorption at 1.6-1.8~\micron, and the level of the $K$-band flux
are all best fit by the average of T7 and T8. Therefore, we measure a 
spectral type of T7.5$\pm0.5$ for HD~3651~B.

\subsection{Physical Properties}
\label{sec:prop}

\subsubsection{Age and Distance}
\label{sec:age}

Now that HN~Peg~B and HD~3651~B have been confirmed as companion T dwarfs,
we examine their physical properties. 
Because these T dwarfs are companions, we can assign to them the ages
and distances of their primaries, which are 
bright, well-studied stars in the solar neighborhood. 
As a result, the ages and distances of these T dwarfs, and hence
the masses, radii, and temperatures, are better constrained than those
of isolated T dwarfs. 
We adopt the {\it Hipparcos} trigonometric distances of $18.4\pm0.3$ and 
$11.1\pm0.1$~pc for HN~Peg~A and HD~3651~A, respectively \citep{per97}.
HN~Peg~A appears to be a young solar analog based on its activity, 
X-ray emission, rotation, Li abundance, and kinematics \citep{gai98}. 
A comparison of the Li measurements for HN~Peg~A \citep{chen01}
to Li data for solar-type members of the Pleiades and Hyades open clusters
and the Ursa Major moving group 
\citep{sod90,sod93a,sod93b} indicates that HN~Peg~A is probably older than
the Pleiades \citep[$\tau=100$-125~Myr,][]{meynet93,stauffer98}
and younger than Ursa Major \citep[$\tau=400$-600~Myr,][]{kin03}
and the Hyades \citep[$\tau=575$-675~Myr,][]{per98}.
A comparison of the rotational properties of HN~Peg~A 
\citep[P=4-5~d, v~sin~i$\sim10$~km~s$^{-1}$, e.g.,][]{kon06,nor04}
and these open clusters \citep{rad87,sod93c,sm93} produces the same result.
The chromospheric activity of HN~Peg~A suggests an age of $\sim0.3$-0.4~Gyr 
\citep{gai98,roc04}.
A debris disk was recently detected around HN~Peg~A at 30 and 70~\micron\ by 
the {\it Spitzer Space Telescope} \citep{bry06,bei06}, which tends to
further support the youth of this star. 
Meanwhile, HD~3651 is probably older than the Sun considering its 
fairly long rotational period \citep[P=48~d,][]{noy84}.
\citet{wri04} estimated an age of 5.9~Gyr from its chromospheric activity 
and \citet{val05} reported an isochronal age of $8.2^{+4.3}_{-5.2}$~Gyr.
Based on these age constraints, we adopt ages of 
$0.3\pm0.2$ and $7\pm3$~Gyr for the HN~Peg and HD~3651 systems, respectively.

\subsubsection{Luminosity}
\label{sec:lbol}

To measure the bolometric luminosities of HN~Peg~B and HD~3651~B, we
first used the $JHK$ photometry to flux calibrate the 0.8-2.5~\micron\ 
spectra. For each object, we constructed a spectral energy distribution
that consisted of these calibrated spectra, the IRAC photometry 
(3.2-9.2~\micron), a linear interpolation of the fluxes between 2.5~\micron\ 
and 3.2~\micron, and a Rayleigh-Jeans distribution longward of 9.2~\micron.
By summing the flux in each distribution and correcting for distance, 
we measured log~$L/L_{\rm \odot}=-4.77\pm0.03$ and $-5.60\pm0.05$
for HN~Peg~B and HD~3651~B, respectively.

\begin{figure}
\plotone{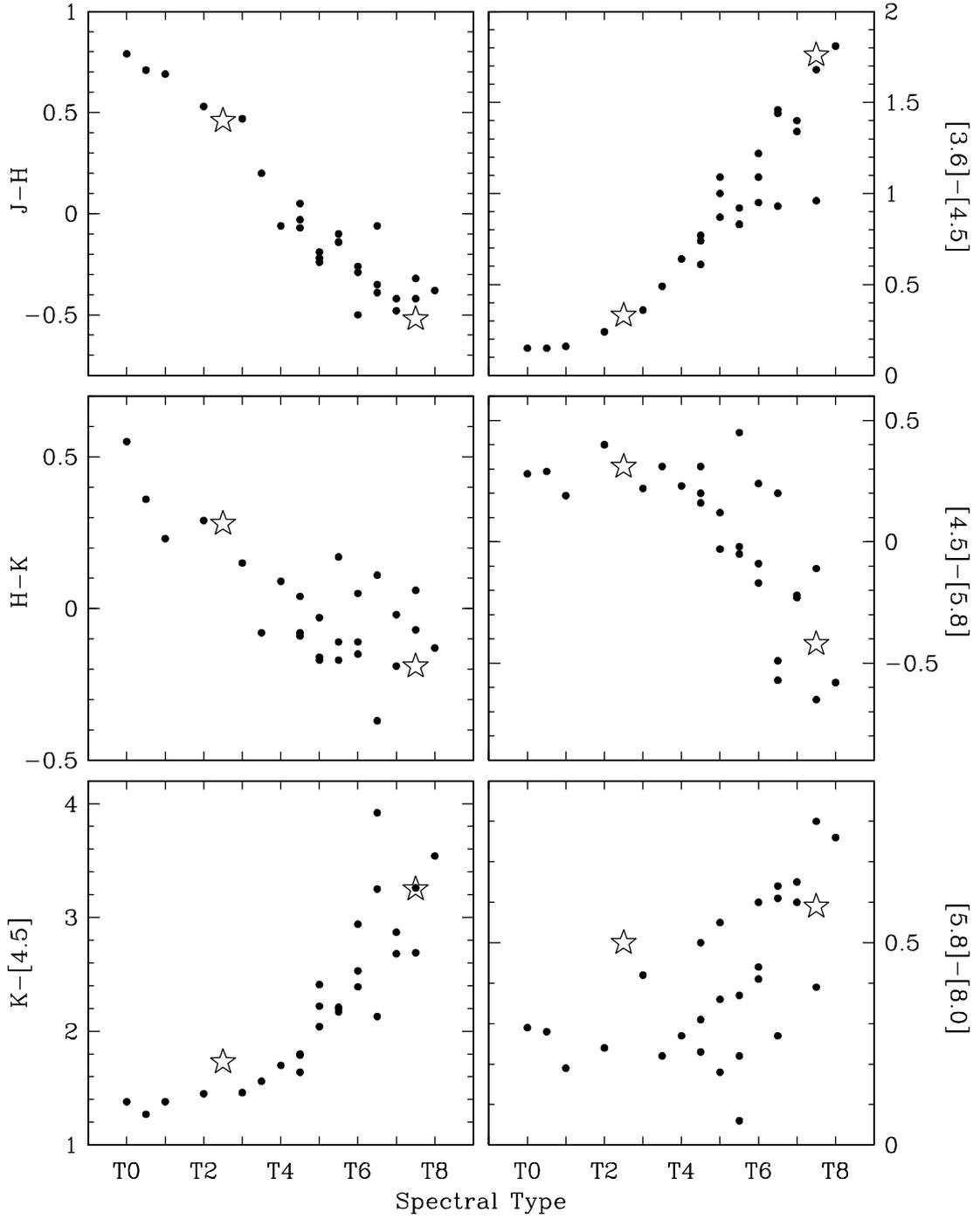}
\caption{ 
Infrared colors versus spectral type for previously known T dwarfs
\citep[{\it filled circles},][]{pat06a} and the new T dwarf companions HN~Peg~B 
({\it left star}) and HD~3651~B ({\it right star}).
}
\label{fig:colors}
\end{figure}

Our luminosity measurement for HN~Peg~B is greater than the value 
produced by combining its $K$ magnitude with the bolometric corrections 
(BC$_K$) that have been previously measured for T2-T3 dwarfs \citep{gol04}. 
For instance, the bolometric luminosity for the T2 dwarf SDSS~1254-0122 
from \citet{cus06} implies BC$_K=3.04$. If we combine this value of 
BC$_K$ with our $K$ measurement for HN~Peg~B and an absolute bolometric 
magnitude for the Sun of $M_{\rm bol \odot}=4.75$, we arrive at 
log~$L/L_{\rm \odot}=-4.83$.
To explore the source of the difference in these two luminosity estimates, 
we compare in Figure~\ref{fig:colors} the near- and mid-IR colors of HN~Peg~B 
to colors of T dwarfs compiled by \citet{pat06a}. The 2MASS measurements 
from \citet{pat06a} have been transformed to the MKO system \citep{sl04}. 
In the diagrams of colors versus spectral type in Figure~\ref{fig:colors}, 
HN~Peg~B is similar to other T dwarfs near its spectral type in terms
of $J-H$, $H-K$, $[3.6]-[4.5]$, and $[4.5]-[5.8]$, but it is redder by 
$\sim0.3$ and $\sim0.15$~mag in $K-[4.5]$ and $[5.8]-[8.0]$, respectively.
The latter difference is not significant because of the large photometric
error in $[8.0]$. However, HN~Peg~B clearly has redder $JHK-$IRAC colors 
relative to typical field dwarfs,
which explains why the luminosity derived from $K$ and BC$_K$ was lower than
the value measured from the photometry and spectra. 
In other words, if we reduced the IRAC fluxes of HN~Peg~B by $\sim30$\% 
so that its $JHK-$IRAC colors were consistent with those of other T2-T3 
dwarfs like SDSS~1254-0122, then our luminosity measurement would agree with 
the one based on $K$ and BC$_K$.

Our luminosity measurement for HD~3651~B agrees well with the value 
produced by combining its $K$ magnitude with BC$_K$ for T7-T8 dwarfs from
\citet{gol04}. In addition, the near- and mid-IR colors of HD~3651~B 
are consistent with those of the other T dwarfs shown in 
Figure~\ref{fig:colors}.

\subsubsection{Mass, Radius, and Temperature}

We use theoretical evolutionary models to convert our luminosity and age
estimates for HN~Peg~B and HD~3651~B to masses, effective temperatures, 
and radii. We perform this conversion by plotting these companions on diagrams 
of luminosity versus age and comparing their positions to lines of 
constant mass, temperature, and radius that are predicted by evolutionary
models. These diagrams are presented in Figs.~\ref{fig:bur97} and 
\ref{fig:bar03} with the models of \citet{bur97} and \citet{bar03}, 
respectively, which imply nearly identical physical properties. 
The models produce $M=0.021\pm0.009$~$M_\odot$, 
$R=0.108^{+0.014}_{-0.006}$~$R_\odot$, and $T_{\rm eff}=1130\pm70$~K 
for HN~Peg~B and $M=0.051\pm0.014$~$M_\odot$, $R=0.082\pm0.006$~$R_\odot$, and 
$T_{\rm eff}=810\pm50$~K for HD~3651~B. 
The spectral types and physical properties of HN~Peg~B and HD~3651~B are
summarized in Table~\ref{tab:prop}.

\section{Discussion}
\label{sec:disc}

We have presented the discovery of two T dwarf companions in the solar
neighborhood, HN~Peg~B (T2.5) and HD~3651~B (T7.5). 
These T dwarfs are the sixth and seventh known
T dwarf companions to stars and are the first T dwarfs discovered with the
{\it Spitzer Space Telescope}. In this section, we describe additional
notable aspects of these new brown dwarfs.

HN~Peg~A has been the target of high-precision radial velocity monitoring 
\citep{cum99,fis05,kon06} and near-IR coronographic imaging \citep{mz04}, 
but no companions have been found through those observations. 
HN~Peg~B was bright enough to be detected by the imaging of \citet{mz04},
but it was outside of their field of view ($r<15\arcsec$). 
This companion was also beyond the field of view of unpublished images
of HN~Peg obtained with the Near-Infrared Camera and Multi-Object Spectrometer
aboard the {\it Hubble Space Telescope} during General Observer program 10176.
Various data from the literature indicate that HN~Peg~A has an age of 
0.1-0.5~Gyr (\S~\ref{sec:age}). 
If the T dwarf S~Ori~70 \citep{zap02} is a field dwarf rather than
a member of the young cluster $\sigma$~Ori \citep{bur04}, 
then HN~Peg~B may be the youngest known T dwarf.
HN~Peg~B closely resembles the recently discovered late-L companion 
to HD~203030 in terms of age and mass \citep{mh06}.

\begin{figure}
\plotone{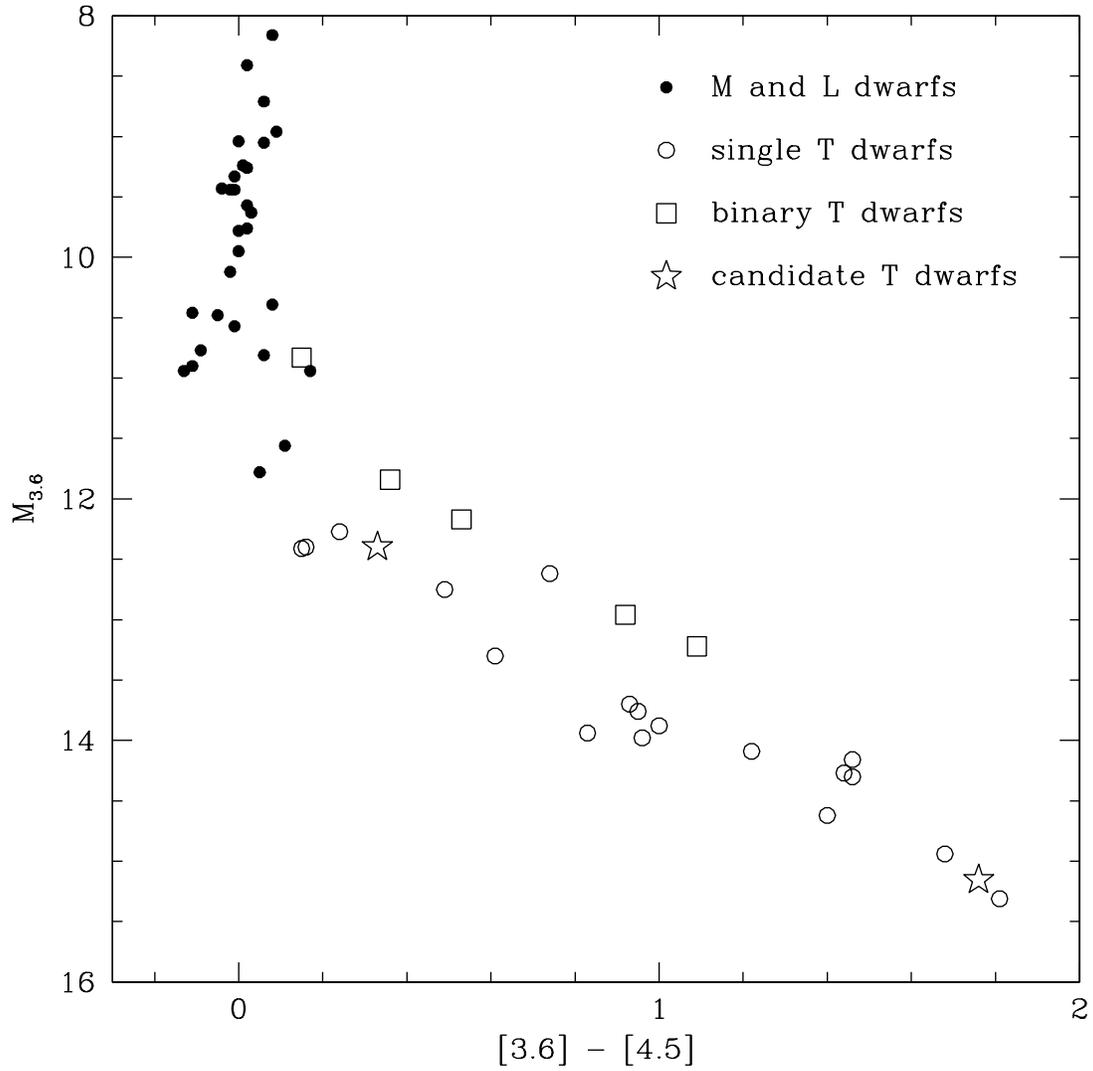}
\caption{ 
IRAC color-magnitude diagram for cool field dwarfs that have measured distances
\citep[{\it open and filled circles and squares},][]{pat06a} and two candidate 
companions of HN~Peg ({\it left star}) and HD~3651 ({\it right star}). 
The absolute magnitudes of the candidates have been computed by adopting 
the distances of the primaries. The magnitudes and colors of these 
candidate companions are indicative of T dwarfs. 
Unlike known T dwarf binaries \citep[][references therein]{bur06ppv},
HN~Peg and HD~3651 are near the lower envelope of the T dwarf
sequence, indicating that they are probably not equal-component binaries.
}
\label{fig:cmd}
\end{figure}

\begin{figure}
\plotone{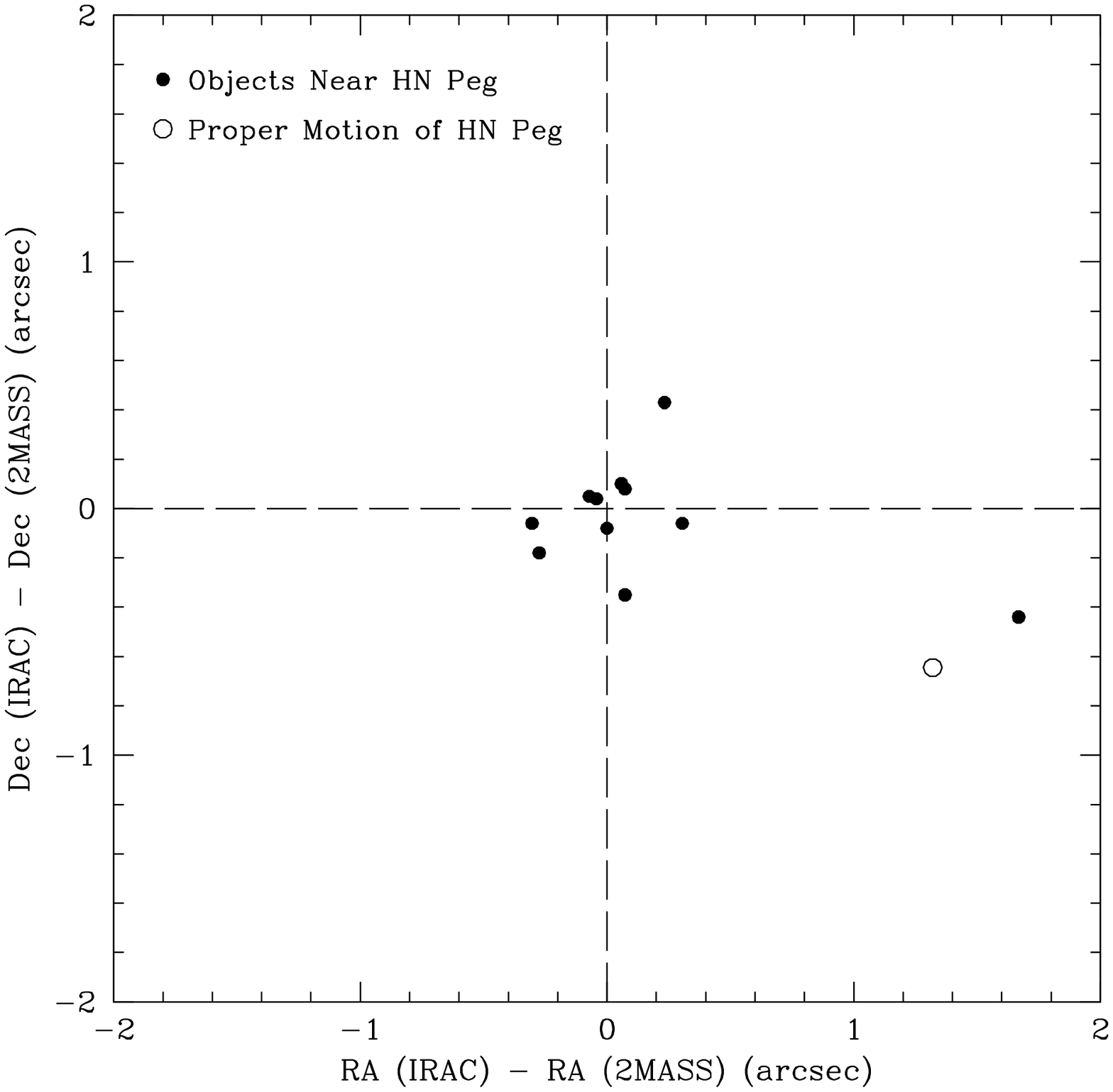}
\caption{
Differences in coordinates of sources near HN~Peg between an $H$-band 2MASS 
image obtained in 1998 and a 4.5~\micron\ IRAC image obtained in 2004 
({\it filled circles}). 
The expected proper motion of HN~Peg during this time interval is indicated
\citep[{\it circle},][]{per97}. 
Excluding the candidate T dwarf companion from Figs.~\ref{fig:cc} and 
\ref{fig:cmd} ({\it rightmost filled circle}), the standard deviations of the 
offsets are $\sigma_{\rm RA}=0\farcs19$ and $\sigma_{\rm Dec}=0\farcs20$,
which represent the errors in these measurements.
The candidate T dwarf exhibits a proper motion that is non-zero at a level
of 8.8~$\sigma_{\rm RA}$ and 2.2~$\sigma_{\rm Dec}$ and is consistent with 
that of HN~Peg at a level of 1.8~$\sigma_{\rm RA}$ and 1.0~$\sigma_{\rm Dec}$,
demonstrating that it is a companion.
}
\label{fig:pm1}
\end{figure}

\begin{figure}
\plotone{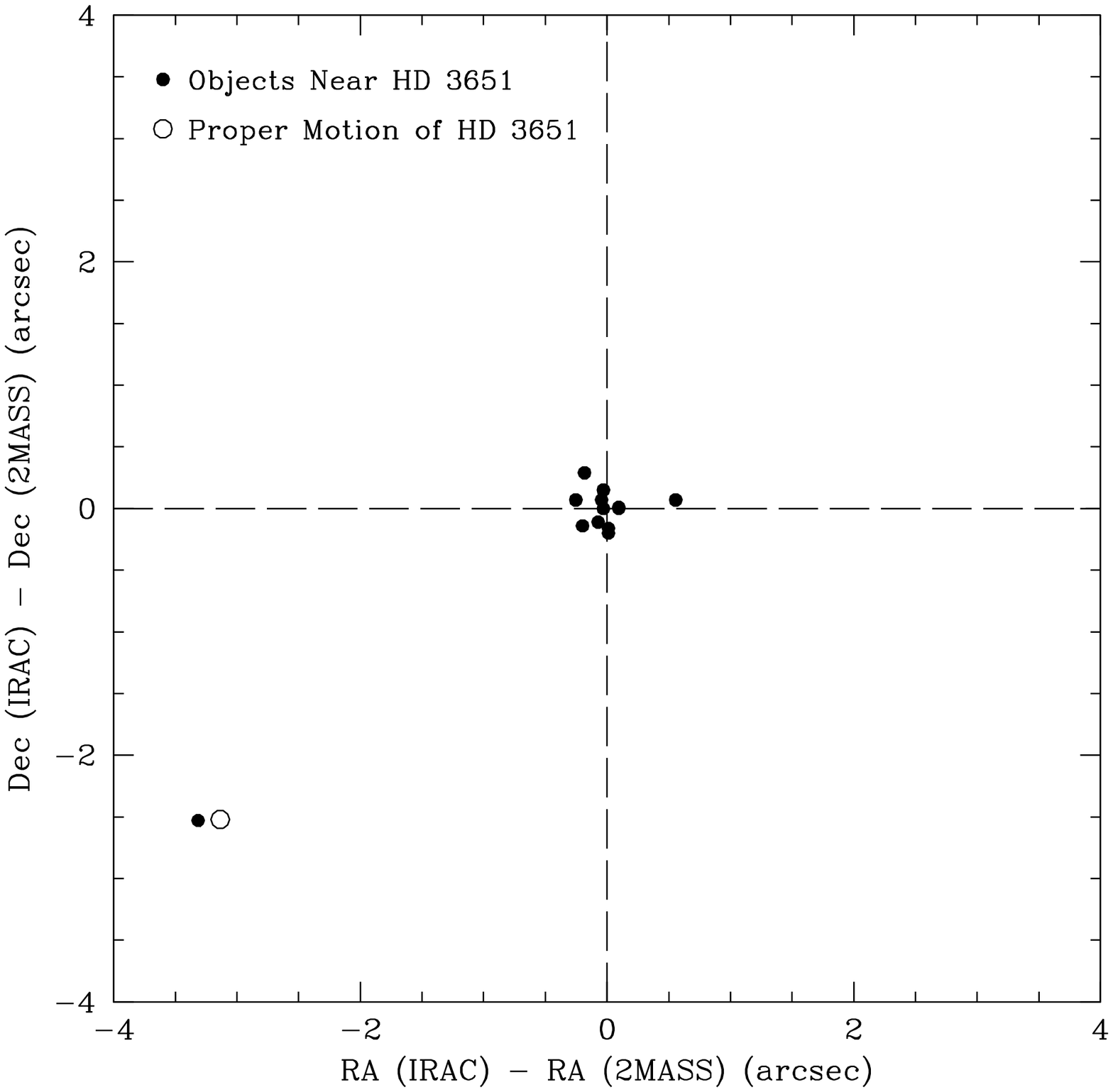}
\caption{
Differences in coordinates of sources near HD~3651 between a $J$-band 2MASS 
image obtained in 1997 and a 4.5~\micron\ IRAC image obtained in 2004 
({\it filled circles}). 
The expected proper motion of HD~3651 during this time interval is indicated
\citep[{\it circle},][]{per97}. 
Excluding the candidate T dwarf companion from Figs.~\ref{fig:cc} and 
\ref{fig:cmd} ({\it leftmost filled circle}), the standard deviations of the 
offsets are $\sigma_{\rm RA}=0\farcs21$ and $\sigma_{\rm Dec}=0\farcs14$,
which represent the errors in these measurements.
The candidate T dwarf exhibits a proper motion that is non-zero at a level
of 16~$\sigma_{\rm RA}$ and 18~$\sigma_{\rm Dec}$ and is consistent with that 
of HD~3651 at a level of 0.8~$\sigma_{\rm RA}$ and 0.1~$\sigma_{\rm Dec}$,
demonstrating that it is a companion.
}
\label{fig:pm2}
\end{figure}

Several properties of HN~Peg~B do not behave in the manner predicted
for young, low-gravity T dwarfs. 
Although the colors within the $JHK$ bands and within the IRAC bands are 
normal, all colors containing one of the former and one of the latter
are redder than those of typical T dwarfs by $\sim0.3$~mag (\S~\ref{sec:lbol}). 
In contrast, low gravity should cause these colors to become bluer, not
redder, according to the theoretical spectra of \citet{bur06}.
Based on its position in color-magnitude diagrams, HN~Peg~B is probably not 
an equal-component binary, but it could be a binary with two unequal 
components\footnote{One of the T dwarfs that has been unresolved in previous 
imaging, 2M 0559-1404 \citep{bur03}, is overluminous in Figure~\ref{fig:cmd}, 
which suggests that it could be a binary. This brown dwarf was identified
as a possible binary because of its bright $J$ magnitude relative to late L 
dwarfs \citep{dah02}, but more recent 1-5~\micron\ measurements
indicated that it might not be significantly overluminous compared to 
other T dwarfs \citep{gol04,vrba04}.}.
Because later T dwarfs emit proportionately more flux in the
mid-IR bands than at shorter wavelengths, an unresolved system of this kind 
(e.g., T1+T5) might be responsible for these anomalous colors. 
Another possible explanation is photometric variability between the dates of
the IRAC and $JHK$ observations \citep{bur02,eno03}. 

\begin{figure}
\plotone{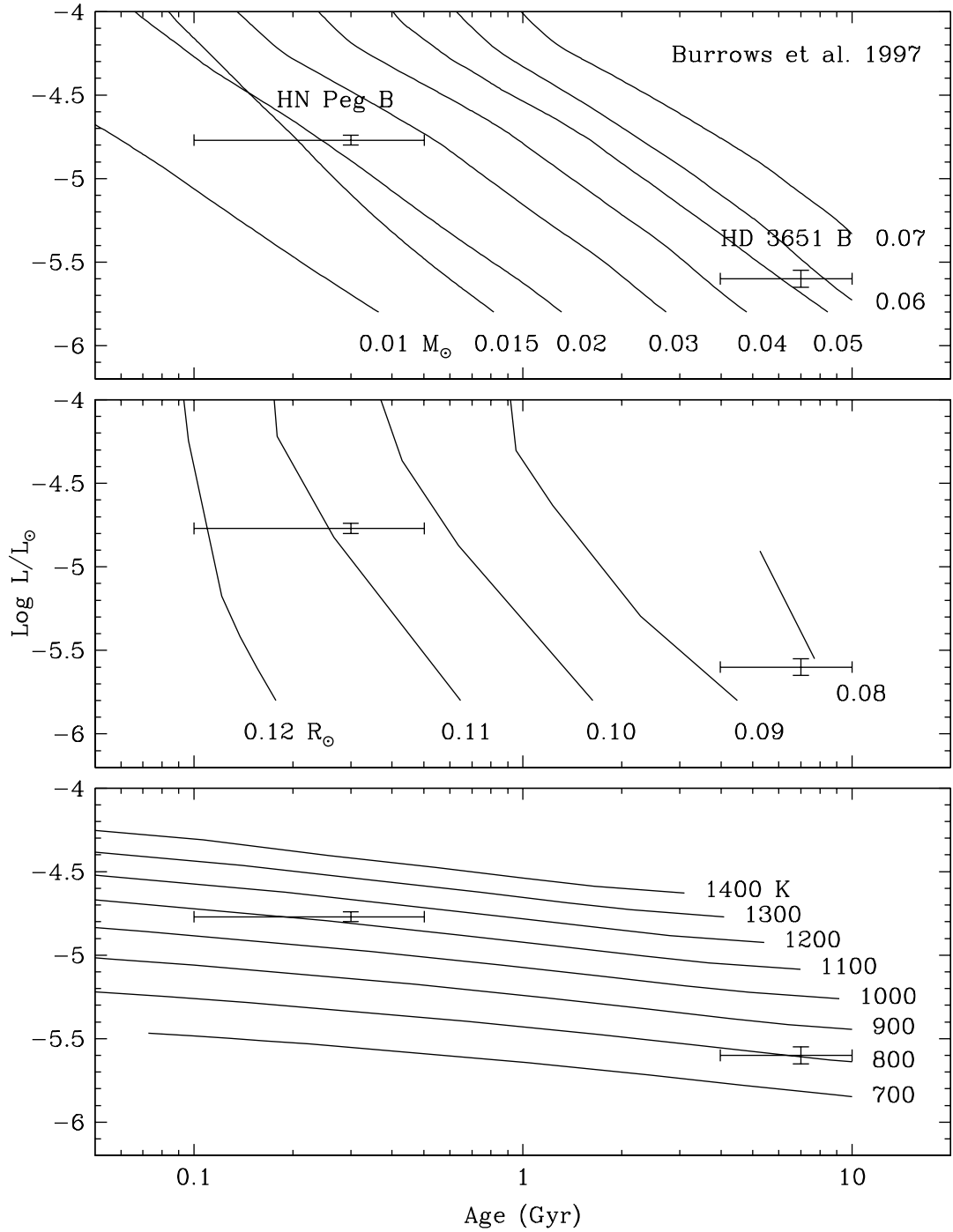}
\caption{ 
Luminosities of HN~Peg~B and HD~3651~B compared to the 
luminosities as a function of age predicted by the theoretical 
evolutionary models of \citet{bur97} ({\it lines}) for constant values 
of mass, radius, and temperature ({\it top to bottom}). 
}
\label{fig:bur97}
\end{figure}

\begin{figure}
\plotone{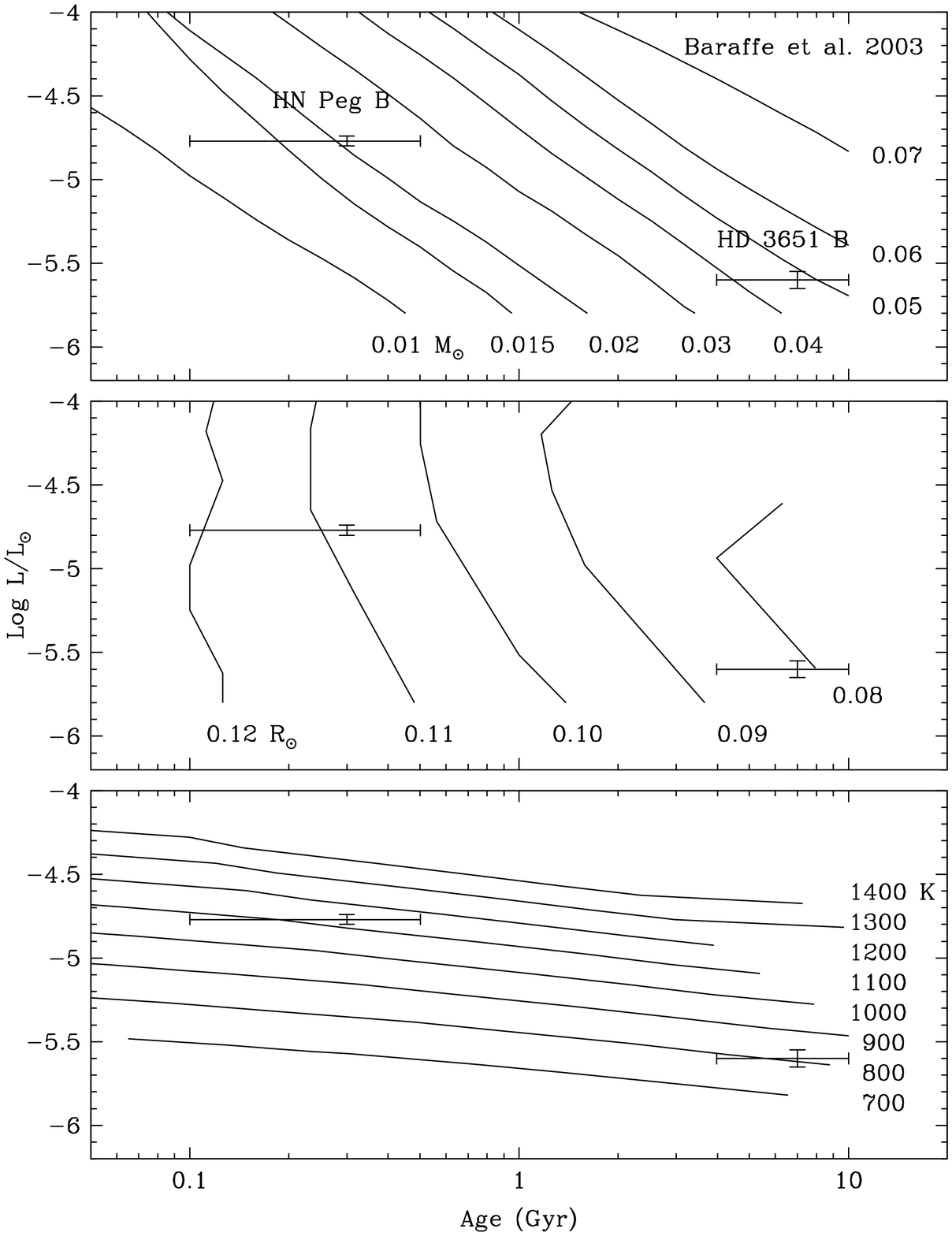}
\caption{ 
Same as Figure~\ref{fig:bur97}, but with the models of \citet{bar03}.
}
\label{fig:bar03}
\end{figure}

Our estimates of the stellar parameters of HN~Peg~B also exhibit peculiarities.
When we compare the luminosity and age of HN~Peg~B to the 
predictions of theoretical evolutionary models, we derive 
an effective temperature of $1130\pm70$~K, which is lower than the values
previously measured for T dwarfs at T2-T3
\citep[$T_{\rm eff}\sim1400$~K,][]{gol04}.
In contrast, the spectral modeling of late T dwarfs by \citet{bur06b}
found that temperature should increase rather than decrease with lower gravity 
at a given spectral type. 
If this prediction is combined with the models shown in Figs.~\ref{fig:bur97}
and \ref{fig:bar03}, then a young T dwarf should be much brighter than 
an older one at the same spectral type. 
Instead, we find that HN~Peg~B is at the faint end of T0-T4 dwarfs 
\citep{gol04} and is at least one magnitude fainter in bolometric luminosity
than expected for its assumed age.
One possible explanation for these discrepancies is that the HN~Peg system is
much older than various data for the primary seem to indicate 
(\S~\ref{sec:age}). However, \citet{mh06} have recently reported the same 
anomalous behavior for a late-L companion near another star that appears 
to be young, HD~203030, and it seems unlikely that the age estimates for 
both stars have such large errors. 
Instead, the presence of this phenomenon in both HN~Peg~B and HD~203030~B 
strongly indicates that the temperature of the L/T transition is 
dependent on surface gravity, and in a manner that has not been previously
predicted. 

When combined with the evolutionary models, the luminosity and age of HN~Peg~B 
translate to a mass $0.021\pm0.009$~$M_\odot$, making it one of the least 
massive known T dwarfs. In addition, 
like HD~203030~B \citep[$M=0.023$~$M_\odot$][]{mh06}, 
HN~Peg~B is among the least massive known brown dwarf companions and 
is a more evolved counterpart to the widely-separated companions
DH~Tau~B \citep[$\tau\sim1$~Myr,][]{ito05}, 
CHXR~73~B \citep[$\tau\sim2$~Myr,][]{luh06cha}, and 
AB~Pic~B \citep[$\tau\sim30$~Myr,][]{cha05b}, which have masses of 
0.01-0.02~$M_\odot$ \citep{luh06cha}.

For HD~3651, radial velocity monitoring has uncovered a sub-Saturn-mass planet 
at 0.3~AU \citep{fis03,fis05} while no companions have been identified in 
adaptive optics images \citep{car05}.
The latter observations did not detect HD~3651~B because their field of 
view was too small ($r<15\arcsec$). 
\citet{tak06} suggested that the high eccentricities of planets in
old, single-planet systems like HD~3651 might be caused by secular 
perturbations from distant low-mass companions, and indeed we have
found such a companion in the form of HD~3651~B. 
This T dwarf is the first resolved substellar companion that has been
discovered in one of the extrasolar planetary systems identified in 
radial velocity surveys. 
Like HN~Peg~B, HD~3651~B is probably not an equal-magnitude binary brown
dwarf according to its position in the color-magnitude diagram
in Figure~\ref{fig:cmd}.
HD~3651~B is a virtual twin to another T dwarf companion, Gl~570~D, in terms 
of spectral type, luminosity, mass, temperature, and radius \citep{bur00,geb01}.
As with Gl~570~D, these measurements have relatively high accuracy because
of the companionship of HD~3651~B to a nearby, well-studied star, 
making it a valuable brown dwarf for calibrating methods of characterizing 
the physical properties of late T dwarfs \citep{bur06a}.

\acknowledgements
We thank Eric Mamajek for discussions regarding the ages of HN~Peg and HD~3651 
and Adam Burgasser for providing his spectra of T dwarf standards.
We are also grateful to Geoff Bryden for checking his 24~\micron\ images 
of HN~Peg for a detection of the companion.
K. L. was supported by grant NAG5-11627 from the NASA Long-Term Space
Astrophysics program. This work is based on observations made
with the {\it Spitzer Space Telescope}, which is 
operated by the Jet Propulsion Laboratory at the California 
Institute of Technology under NASA contract 1407.
Support for the IRAC instrument was provided by NASA through contract
960541 issued by JPL.

\clearpage

\begin{deluxetable}{llllllllll}
\tabletypesize{\scriptsize}
\tablewidth{0pt}
\tablecaption{Astrometry and Photometry for Companions\label{tab:data}}
\tablehead{
\colhead{} &
\colhead{$\rho$} &
\colhead{PA} &
\colhead{} &
\colhead{} &
\colhead{} &
\colhead{} &
\colhead{} &
\colhead{} &
\colhead{} \\
\colhead{Companion} &
\colhead{(arcsec)} &
\colhead{(deg)} &
\colhead{$J$(MKO)} &
\colhead{$H$(MKO)} &
\colhead{$K$(MKO)} &
\colhead{$[3.6]$} &
\colhead{$[4.5]$} &
\colhead{$[5.8]$} &
\colhead{$[8.0]$} 
}
\startdata
HN Peg B & 43.2$\pm$0.4 & 254.4$\pm$0.6 & 15.86$\pm$0.03 & 15.40$\pm$0.03 & 15.12$\pm$0.03 & 13.72$\pm$0.04 & 13.39$\pm$0.02 & 13.08$\pm$0.10 & 12.58$\pm$0.11 \\
HD 3651 B & 42.9$\pm$0.4 & 290.1$\pm$0.6 & 16.16$\pm$0.03 & 16.68$\pm$0.04 & 16.87$\pm$0.05 & 15.38$\pm$0.04 & 13.62$\pm$0.02 & 14.04$\pm$0.12 & 13.45$\pm$0.14 \\
\enddata
\end{deluxetable}

\begin{deluxetable}{lllllllll}
\tabletypesize{\scriptsize}
\tablewidth{0pt}
\tablecaption{Properties of Companions\label{tab:prop}}
\tablehead{
\colhead{} &
\colhead{distance\tablenotemark{a}} &
\colhead{$\rho$\tablenotemark{b}} &
\colhead{Spectral} &
\colhead{Age} &
\colhead{} &
\colhead{} &
\colhead{} &
\colhead{$T_{\rm eff}$\tablenotemark{c}} \\
\colhead{Companion} &
\colhead{(pc)} &
\colhead{(AU)} &
\colhead{Type} &
\colhead{(Gyr)} &
\colhead{log $L/L_\odot$} &
\colhead{$M/M_\odot$\tablenotemark{c}} &
\colhead{$R/R_\odot$\tablenotemark{c}} &
\colhead{(K)} 
}
\startdata
HN Peg B & 18.4$\pm$0.3 & 795$\pm$15 & T2.5$\pm$0.5 & 0.3$\pm$0.2 & $-$4.77$\pm0.03$ 
& 0.021$\pm$0.009 & 0.108$^{+0.014}_{-0.006}$ & 1130$\pm$70 \\
HD 3651 B & 11.1$\pm$0.1 & 476$\pm$6 & T7.5$\pm$0.5 & 7$\pm$3 & $-$5.60$\pm0.05$ & 0.051$\pm$0.014 & 0.082$\pm$0.006 & 810$\pm$50 \\
\enddata
\tablenotetext{a}{Distance of primary \citep{per97}.}
\tablenotetext{b}{Projected physical separation from primary.}
\tablenotetext{c}{Estimated from the age, log~$L/L_\odot$, and the models of 
\citet{bur97} and \citet{bar03} in Figs.~\ref{fig:bur97} and \ref{fig:bar03}.}
\end{deluxetable}

\begin{thebibliography}{}

\bibitem[Allard et al.(2001)]{all01}
Allard, F., Hauschildt, P. H., Alexander, D. R., Tamanai, A., \& Schweitzer, A.
2001, \apj, 556, 357

\bibitem[Baraffe et al.(2003)]{bar03}
Baraffe, I., Chabrier, G., Barman, T. S., Allard, F., \& Hauschildt, P. H.
2003, \aap, 402, 701

\bibitem[Beichman et al.(2006)]{bei06}
Beichman, C. A., et al. 2006, \apj, 639, 1166

\bibitem[Biller et al.(2006)]{bil06}
Biller, B. A., Kasper, M., Close, L. M., Brandner, W., \& Kellner, S. 2006, 
\apj, 641, L141




\bibitem[Bryden et al.(2006)]{bry06}
Bryden, G., et al. 2006, \apj, 636, 1098

\bibitem[Burgasser et al.(2006a)]{bur06a}
Burgasser, A. J., Burrows, A., \& Kirkpatrick, J. D. 2006a, \apj, 639, 1095

\bibitem[Burgasser et al.(2006b)]{bur06b}
Burgasser, A. J., Geballe, T. R., Leggett, S. K., Kirkpatrick, J. D., 
\& Golimowski, D. A. 2006b, \apj, 637, 1067

\bibitem[Burgasser et al.(2005)]{bur05}
Burgasser, A. J., Kirkpatrick, J. D., \& Lowrance, P. J. 2005, \aj, 129, 2849 

\bibitem[Burgasser et al.(2004)]{bur04}
Burgasser A. J., Kirkpatrick J. D., McGovern M. R., McLean I. S., Prato L., 
\& Reid, I. N. 2004, \apj, 604, 827

\bibitem[Burgasser et al.(2003)]{bur03}
Burgasser, A. J., Kirkpatrick, J. D., Reid, I. N., Brown, M. E., Miskey, C. L., 
\& Gizis, J. E. 2003, \apj, 586, 512 

\bibitem[Burgasser et al.(2006c)]{bur06ppv}
Burgasser, A. J., Reid, I. N., Siegler, N., Close, L., Allen, P., Lowrance, P.,
Gizis, J. 2006c, Protostars and Planets V, in press

\bibitem[Burgasser et al.(2000)]{bur00}
Burgasser, A. J., et al. 2000, \apj, 531, L57

\bibitem[Burgasser et al.(2002)]{bur02}
Burgasser, A. J., et al. 2002, \apj, 571, L151

\bibitem[Burrows et al.(2006)]{bur06}
Burrows, A., Sudarsky, D., Hubeny, I. 2006, \apj, 640, 1063

\bibitem[Burrows et al.(1997)]{bur97}
Burrows, A., et al. 1997, \apj, 491, 856

\bibitem[Carson et al.(2005)]{car05}
Carson, J. C., Eikenberry, S. S., Brandl, B. R., Wilson, J. C., \& Hayward, 
T. L. 2005, \aj, 130, 1212

\bibitem[Chauvin et al.(2003)]{cha03}
Chauvin, G., et al. 2003, \aap, 404, 157

\bibitem[Chauvin et al.(2004)]{cha04}
Chauvin, G., et al. 2004, \aap, 425, L29

\bibitem[Chauvin et al.(2005a)]{cha05a}
Chauvin, G., et al. 2005a, \aap, 438, L25

\bibitem[Chauvin et al.(2005b)]{cha05b}
Chauvin, G., et al. 2005b, \aap, 438, L29

\bibitem[Chauvin et al.(2006)]{chau06}
Chauvin, G., et al. 2006, \aap, in press

\bibitem[Chen et al.(2001)]{chen01}
Chen, Y. Q., Nissen, P. E., Benoni T., \& Zhao, G. 2001, \aap, 371, 943


\bibitem[Cummings et al.(1999)]{cum99}
Cumming, A., Marcy, G. W., \& Butler, R. P. 1999, \apj, 526, 890

\bibitem[Cushing et al.(2004)]{cus04}
Cushing, M. C., Vacca, W. D., \& Rayner, J. T. 2004, \pasp, 116, 362

\bibitem[Cushing et al.(2006)]{cus06}
Cushing, M. C., et al. 2006, \apj, 648, 614

\bibitem[Dahn et al.(2002)]{dah02}
Dahn, C. C., et al. 2002, \aj, 124, 1170 

\bibitem[Els et al.(2001)]{els01}
Els, S.G., Sterzik, M.F., Marchis, F., Pantin, E., Endl, M., \& K\"urster, M.
2001, \aap, 370, L1

\bibitem[Enoch et al.(2003)]{eno03}
Enoch, M. L., Brown, M. E., \& Burgasser, A. J. 2003, \aj, 126, 1006

\bibitem[Fazio et al.(2004)]{faz04}
Fazio, G. G., et al. 2004, \apjs, 154, 10

\bibitem[Fischer et al.(2003)]{fis03}
Fischer, D. A., Butler, R. P., Marcy, G. W., Vogt, S. S., \& Henry, G. W. 2003, 
\apj, 590, 1081

\bibitem[Fischer \& Valenti(2005)]{fis05}
Fischer, D. A., \& Valenti, J. 2005, \apj, 622, 1102

\bibitem[Fix \& Hodge(1951)]{fix51}
Fix, E., \& Hodge, J. L. 1951, Technical Report 21-49-004 4, US Air Force
School of Aviation Medicine, Randolph Field, TX

\bibitem[Gaidos(1998)]{gai98}
Gaidos, E. J. 1998, \pasp, 110, 1259

\bibitem[Geballe et al.(2001)]{geb01}
Geballe, T. R., Saumon, D., Leggett, S. K., Knapp, G. R., Marley, M. S., \& 
Lodders, K. 2001, \apj, 556, 373

\bibitem[Golimowski et al.(2004)]{gol04}
Golimowski, D. A., et al. 2004, \aj, 127, 3516

\bibitem[Guenther et al.(2001)]{gue01}
Guenther, E. W., Neuh\"{a}user, R., Hu\'{e}lamo, N., Brandner, W., \& Alves, J.
2001, \aap, 365, 514


\bibitem[Itoh et al.(2005)]{ito05}
Itoh, Y., et al. 2005, \apj, 620, 984

\bibitem[King et al.(2003)]{kin03}
King, J. R., Villarreal, A. R., Soderblom, D. R., Gulliver, A. F., \&
Adelman, S. J. 2003, \aj, 125, 1980

\bibitem[Kirkpatrick(2005)]{kir05}
Kirkpatrick, J. D. 2005, \araa, 43, 195

\bibitem[Kirkpatrick et al.(2006)]{kir06}
Kirkpatrick, J. D., et al. 2006, \apj, 639, 1120

\bibitem[Knapp et al.(2004)]{kna04}
Knapp, G. R., et al. 2004, \aj, 127, 3553

\bibitem[K\"onig et al.(2006)]{kon06}
K\"onig, B., Guenther, E. W., Esposito, M., \& Hatzes, A. 2006, \mnras, 365,
1050

\bibitem[Lagrange et al.(2006)]{lag06}
Lagrange, A.-M., Beust, H., Udry, S., Chauvin, G., \& Mayor, M. 2006, \aap,
in press

\bibitem[Leggett et al.(2006)]{leg06}
Leggett, S. K., et al. 2006, \mnras, submitted


\bibitem[Lowrance et al.(1999)]{low99}
Lowrance, P. J., et al. 1999, \apj, 512, L69

\bibitem[Lowrance et al.(2000)]{low00}
Lowrance, P. J., et al. 2000, \apj, 541, 390

\bibitem[Lowrance et al.(2005)]{low05} 
Lowrance, P. J., et al. 2005, \aj, 130, 1845





\bibitem[Luhman \& Jayawardhana(2002)]{luh02}
Luhman, K. L., \& Jayawardhana, R. 2002, \apj, 566, 1132

\bibitem[Luhman et al.(2006)]{luh06cha}
Luhman, K. L., et al. 2006, \apj, in press











\bibitem[Marengo et al.(2006)]{mar06}
Marengo, M., Megeath, S. T., Fazio, G. G., Stapelfeldt, K. R., Werner, M. W., 
Backman, D. E. 2006, \apj, 647, 1437





\bibitem[McCarthy \& Zuckerman(2004)]{mz04}
McCarthy, C., \& Zuckerman, B. 2004, \aj, 127, 2871

\bibitem[McCaughrean et al.(2004)]{mcc04}
McCaughrean, M. J., et al. 2004, \aap, 413, 1029

\bibitem[McGovern et al.(2004)]{mc04}
McGovern, M. R., Kirkpatrick, J. D., McLean, I. S., Burgasser, A. J.,
Prato, L., \& Lowrance, P. J. 2004, \apj, 600, 1020

\bibitem[Metchev \& Hillenbrand(2004)]{mh04}
Metchev, S. A., \& Hillenbrand, L. A. 2004, \apj, 617, 1330

\bibitem[Metchev \& Hillenbrand(2006)]{mh06}
Metchev, S. A., \& Hillenbrand, L. A. 2006, \apj, in press

\bibitem[Meynet et al.(1993)]{meynet93}
Meynet, G., Mermilliod, J.-C., \& Maeder, A.  1993, \aaps, 98, 477

\bibitem[Mugrauer \& Neuh\"{a}user(2005)]{mug05a}
Mugrauer, M., \& Neuh\"{a}user, R. 2005, \mnras, 361, L15

\bibitem[Mugrauer et al.(2005)]{mug05b}
Mugrauer, M., Neuh\"{a}user, R., Seifahrt, A., Mazeh, T., \& Guenther, E.
2005, \aap, 440, 1051

\bibitem[Mugrauer et al.(2006)]{mug06}
Mugrauer, M., Seifahrt, A., Neuh\"{a}user, R., \& Mazeh, T. 2006, \mnras,
in press

\bibitem[Nakajima et al.(1995)]{nak95}
Nakajima, T., Oppenheimer, B. R., Kulkarni, S. R., Golimowski, D. A., 
Matthews, K., \& Durrance, S. T. 1995, \nat, 378, 463


\bibitem[Neuh\"{a}user et al.(2000a)]{neu00a}
Neuh\"{a}user, R., et al. 2000a, \aap, 354, L9

\bibitem[Neuh\"{a}user et al.(2000b)]{neu00b}
Neuh\"{a}user, R., et al. 2000b, \aap, 360, L39


\bibitem[Nordstrom et al.(2004)]{nor04}
Nordstr{\"o}m, B., et al. 2004, \aap, 418, 989

\bibitem[Noyes et al.(1984)]{noy84}
Noyes, R. W., Hartmann, L. W., Baliunas, S. L., Duncan, D. K., \& Vaughan, 
A. H. 1984, \apj, 279, 763

\bibitem[Oppenheimer et al.(1995)]{opp95}
Oppenheimer, B. R., Kulkarni, S. R., Nakajima, T., \& Matthews, K. 1995, 
Science, 270, 1478

\bibitem[Patience et al.(2002)]{pat02}
Patience, J., et al. 2002, \apj, 581, 654

\bibitem[Patten et al.(2006a)]{pat06a}
Patten, B. M., et al. 2006a, \apj, in press

\bibitem[Patten et al.(2006b)]{pat06b}
Patten, B. M., et al. 2006b, Protostars and Planets V, in press

\bibitem[Perryman et al.(1997)]{per97}
Perryman, M. A. C., et al. 1997, \aap, 323, L49

\bibitem[Perryman et al.(1998)]{per98}
Perryman, M. A. C., et al. 1998, \aap, 331, 81



\bibitem[Radick et al.(1987)]{rad87}
Radick, Richard R., Thompson, D. T., Lockwood, G. W., Duncan, D. K., \&
Baggett, W. E. 1987, \apj, 321, 459

\bibitem[Rayner et al.(2003)]{ray03}
Rayner, J. T., et al. 2003, \pasp, 115, 362


\bibitem[Rebolo et al.(1998)]{reb98}
Rebolo, R., et al. 1998, Science, 282, 1309   


\bibitem[Rocha-Pinto et al.(2004)]{roc04}
Rocha-Pinto, H. J., Flynn, C., Scalo, J., H\"anninen, J., Maciel, W. J., \& 
Hensler, G. 2004, \aap, 423, 517


\bibitem[Scholz et al.(2003)]{sch03}
Scholz, R.-D., McCaughrean, M. J., Lodieu, N., \& Kuhlbrodt, B. 2003, \aap, 
398, L29

\bibitem[Schuster et al.(2006)]{sch06}
Schuster, M., Marengo, M., \& Patten, B. 2006, Proceedings of the SPIE, 6270-74

\bibitem[Simons \& Tokunaga(2002)]{sim02}
Simons, D. A., \& Tokunaga, A. 2002, \pasp, 114, 169


\bibitem[Soderblom et al.(1993a)]{sod93a}
Soderblom, D. R., Jones, B. F., Balachandran, S., Stauffer, J. R., Duncan,
D. K., Fedele, S. B., \& Hudon, J. D. 1993a, 106, 1059

\bibitem[Soderblom \& Mayor(1993)]{sm93}
Soderblom, D. R., \& Mayor, M. 1993, \apj, 402, 5

\bibitem[Soderblom et al.(1990)]{sod90}
Soderblom, D. R., Oey, M. S., Johnson, D. R. H., \& Stone, R. P. S. 1990, \aj, 
99, 595

\bibitem[Soderblom et al.(1993b)]{sod93b}
Soderblom, D. R., Pilachowski, C. A., Fedele, S. B., \& Jones, B. F. 1993b, 
\aj, 105, 2299

\bibitem[Soderblom et al.(1993c)]{sod93c}
Soderblom, D. R., Stauffer, J. R., Hudon, J. D., \& Jones, B. F. 1993c, \apjs,
85, 315


\bibitem[Stauffer et al.(1998)]{stauffer98}
Stauffer, J. R., Schultz, G., \& Kirkpatrick, J. D. 1998, \apj, 499, 199

\bibitem[Stephens \& Leggett(2004)]{sl04}
Stephens, D. C., \& Leggett, S. K. 2004, \pasp, 116, 9

\bibitem[Takeda et al.(2006)]{tak06}
Takeda, G., Ford, E. B., Sills, A., Rasio, F. A., Fischer, D. A., \& Valenti, 
J. A. 2006, \apjs, in press

\bibitem[Tokunaga et al.(2002)]{tok02}
Tokunaga, A. T., Simons, D. A., \& Vacca, W. D. 2002, \pasp, 114, 180

\bibitem[Tokunaga \& Vacca(2005)]{tok05}
Tokunaga, A. T., \& Vacca, W. D. 2005, \pasp, 117, 421

\bibitem[Vacca et al.(2003)]{vac03}
Vacca, W. D., Cushing, M. C., \& Rayner J. T., 2003, \pasp, 115, 389

\bibitem[Valenti \& Fischer(2005)]{val05}
Valenti, J. A., \& Fischer, D. A. 2005, \apjs, 159, 141

\bibitem[Vrba et al.(2004)]{vrba04}
Vrba, F. J., et al. 2004, \aj, 127, 2948

\bibitem[Werner et al.(2004)]{wer04}
Werner, M. W., et al. 2004, \apjs, 154, 1


\bibitem[Wilson et al.(2001)]{wil01}
Wilson, J. C., Kirkpatrick, J. D., Gizis, J. E., Skrutskie, M. F.,
Monet, D. G., \& Houck, J. R. 2001, \aj, 122, 1989


\bibitem[Wright et al.(2004)]{wri04}
Wright, J. T., Marcy, G. W., Butler, R. P., \& Vogt, S. S. 2004, \apjs, 152, 261

\bibitem[Zapatero Osorio et al.(2002)]{zap02}
Zapatero Osorio, M. R., et al. 2002, \apj, 578, 536

\end{thebibliography}
\end{document}